\RequirePackage{lineno}
\documentclass[aps,twocolumn,showpacs,preprintnumbers,amsmath]{revtex4}
\usepackage{graphicx}% Include figure files
\usepackage{dcolumn}% Align table columns on decimal point
\usepackage{epstopdf}
\usepackage{color}
\usepackage{comment}

\usepackage{bm}% bold math
\begin{document}
	%\linenumbers
	%\begin{frontmatter}
	\title{Deconfinement and degrees of freedom in $pp$ and $A-A$ collisions at LHC energies}
	\author{Aditya N Mishra{$^{1,2}$},  Guy Pai{\'c}{$^1$}, C. Pajares{$^3$}, R. P. Scharenberg{$^4$ }  and B. K. Srivastava{$^4$}}
	\medskip
	\affiliation{$^1$Instituto de Ciencias Nucleares, Universidad Nacional Aut\'onoma de M\'exico, Apartado Postal 70-543,
		M\'exico Distrito Federal 04510, M\'exico\\
		$^2$Wigner Research Center for Physics, H-1121Budapest, Hungary\\
		$^3$Departamento de Fisica de Particulas, Universidale de Santiago de Compostela and Instituto Galego de Fisica de Atlas Enerxias(IGFAE), 15782 Santiago, de Compostela, Spain \\
		$^4$Department of Physics and Astronomy, Purdue University, West Lafayette, IN-47907, USA}
	\bigskip
	\date{\today}% It is always \today, today      

\begin{abstract}
We present the extraction of the temperature by analyzing the charged particle transverse momentum spectra in lead-lead (Pb-Pb) and proton-proton (${\bf pp}$) collisions at LHC energies from the ALICE Collaboration using the Color String Percolation Model (CSPM). From the measured  energy density ${\bm \varepsilon}$ and the temperature  T the  dimensionless quantity ${\bm \varepsilon/}T^{4}$ is  obtained to get the degrees of freedom (DOF),  ${\bm \varepsilon}/T^{4}$ = DOF ${ \pi^{2}}$/30.
  We observe for the first time a two-step behavior in the increase of  DOF, characteristic of deconfinement, above the hadronization temperature  at temperature $\sim$ 210 MeV for both Pb-Pb and  ${\bf pp}$ collisions and a sudden increase to the ideal gas value of  $\sim $ 47 corresponding to three quark flavors in the case of Pb-Pb collisions.
\end{abstract}

\maketitle
\section{Introduction}
\label{intro}
The Quantum Chromodynamics (QCD) phase diagram is closely related to the history of the universe and can be probed by heavy ion collisions. Of particular interest in the heavy ion collision experiments are the details of the deconfinement and chiral transitions which determine the QCD phase diagram.  One of the main challenges of the field is to simultaneously determine the temperature and the energy density of the matter produced in a collision and hence the number of thermodynamic degrees of freedom (DOF) \cite{busza}.
     The present work explores the initial stage of high energy  collisions at LHC energies analyzing the  published ALICE data \cite{alicepp,alicepb} on the transverse momentum ($p_{t}$) spectra of charged particles using the framework of the clustering of color sources (CSPM) \cite{review15}.
This approach has been successfully used to describe the initial stages in the soft region in high energy nucleus-nucleus and nucleon-nucleon collisions \cite{review15,eos,eos2,IS2013,eos3,pp19}. The CSPM is in fact different from the hydrodynamics picture and is more in line with other studies where the interaction among strings \cite{bierlich,bierlich2,ortiz} or the domain color structure \cite{muller,lappi} is taken into account.
The determination of  the DOF requires the  measurement of the initial thermalized (maximum entropy) temperature and the initial energy density at time $\sim 1$ fm/c of the hot matter produced in high energy A-A and  $pp$ collisions. Lattice Quantum Chromo Dynamics simulations (LQCD) indicate that the non-perturbative region of hot QCD matter extends up to temperature of 400 MeV, well above the universal hadronization temperature \cite{latthigh}.

\section{Clustering of color sources and percolation}

Multiparticle production is currently described in terms of color strings stretched between the projectile and the target, which decay into new strings and subsequently hadronize to produce observed hadrons. Color strings may be viewed as small areas in the transverse plane filled with color field created by colliding partons. With growing energy and size of the colliding system, the number of strings grows, and they start to overlap, forming clusters, in the transverse plane very much similar to disks in two dimensional percolation theory \cite{inch}. At a certain critical density $\xi_{c} \sim $ 1.2 a macroscopic cluster appears that marks the percolation phase transition. For nuclear collisions, this density corresponds to $\xi$ = $N_{s}\frac {S_{1}}{S_{A}}$ where $N_{s}$ is the total number of strings created in the collision, each one of area $S_{1} = \pi r_{0}^{2}$ and $S_{A}$ corresponds to nuclear overlap area, with $r_{0} \approx$ 0.2 fm.  
This is the Color String Percolation Model (CSPM) \cite{pajares1,pajares2}.
  
The interaction between strings occurs forming clusters when they overlap and the general result, due to the SU(3) random summation of charges, is a reduction in multiplicity and an increase in the average transverse momentum squared, $\langle p_{t}^{2} \rangle$.  The strings decay into new ones through color neutral  $q-\bar{q}$ pairs production. The Schwinger QED$_{2}$ string breaking mechanism  produces these   $q-\bar{q}$ pairs at time $\tau_{pro} \sim $ 1 fm/c, which  subsequently hadronize to  produce the observed hadrons \cite{wong}. Schwinger mechanism has also been used in the decay of color flux tubes produced by the quark-gluon plasma for modeling the initial  stages in heavy ion collisions \cite{prc1,prc2}. 

The combination of  the string density dependent cluster formation and the 2D percolation clustering phase transition, are the basic elements of the non-perturbative CSPM. The percolation theory governs the geometrical pattern of string clustering. Its observable implications, however, require the introduction of some dynamics in order to describe the behavior of the cluster formed by several overlapping clusters. 
We assume that a cluster of ${\it n}$ strings that occupies an area of $S_{n}$ behaves as a single color source with a higher color field $\vec{Q_{n}}$ corresponding to the vectorial sum of the color charges of each individual string $\vec{Q_{1}}$. The resulting color field covers the area of the cluster. As $\vec{Q_{n}} = \sum_{1}^{n}\vec{Q_{1}}$, and the individual string colors may be oriented in an arbitrary manner respective to each other , the average $\vec{Q_{1i}}\vec{Q_{1j}}$ is zero, and $\vec{Q_{n}^2} = n \vec{Q_{1}^2} $.
Knowing the color charge, one can compute the multiplicity $\mu_{n}$ and the mean transverse momentum squared $\langle p_{t}^{2} \rangle_{n}$ of the particles produced by a cluster, which are proportional to the color charge and color field, respectively \cite{pajares2}
\begin{equation}
\mu_{n} = \sqrt {\frac {n S_{n}}{S_{1}}}\mu_{0};\hspace{5mm}
%\end{equation}
%\begin{linenomath*}
 %\begin{equation}                                      
\langle p_{t}^{2} \rangle_{n} = \sqrt {\frac {n S_{1}}{S_{n}}} {\langle p_{t}^{2} \rangle_{1}},
\label{mu}
\end{equation}
where $\mu_{0}$ and $\langle p_{t}^{2}\rangle_{1}$ are the mean multiplicity and transverse momentum squared of particles produced from a single string with a transverse area  $S_{1} = \pi r_{0}^2$ with $r_{0}$ = 0.2 fm \cite{review15}.  For strings  just touching each other $S_{n} = n S_{1}$, and $\mu_{n} = n \mu_{0}$, $\langle p_{t}^{2}\rangle_{n}= \langle p_{t}^{2}\rangle_{1}$. When strings fully overlap, $S_{n} = S_{1}$  and therefore 
 $\mu_{n} = \sqrt{n} \mu_{0}$ and $\langle p_{t}^{2}\rangle_{n}= \sqrt{n} \langle p_{t}^{2}\rangle_{1}$, so that the multiplicity is maximally suppressed and the $\langle p_{t}^{2}\rangle_{n}$ is maximally enhanced. This implies a simple relation between the multiplicity and transverse momentum $\mu_{n}\langle p_{t}^{2}\rangle_{n}=n\mu_{0}\langle p_{t}^{2}\rangle_{1}$, which means conservation of the total transverse momentum produced.
 In the thermodynamic limit, one obtains the average value of $ n S_{1}/S_{n}$ for all the clusters \cite{pajares1,pajares2}
 \begin{equation}
\langle \frac {n S_{1}}{S_{n}} \rangle = \frac {\xi}{1-e^{-\xi}}\equiv \frac {1}{F(\xi)^2}
\end{equation}
where $F(\xi)$ is the color suppression factor by which the overlapping strings reduce the net color charge of the strings. With $F(\xi)\rightarrow 1$ as $\xi \rightarrow 0$ and $F(\xi)\rightarrow 0$ as $\xi \rightarrow \infty $, where  $\xi = \frac {N_{s} S_{1}}{S_{N}}$ is the percolation density parameter.
Eq.~(\ref{mu}) can be written as $\mu_{n}=n F(\xi)\mu_{0}$ and 
$\langle p_{t}^{2}\rangle_{n} ={\langle p_{t}^{2} \rangle_{1}}/F(\xi)$.  
It is worth noting that CSPM is a saturation model similar to the Color Glass Condensate (CGC),
where $ {\langle p_{t}^{2} \rangle_{1}}/F(\xi)$ plays the same role as the saturation momentum scale $Q_{s}^{2}$ in the CGC model \cite{cgc,perx}.
\begin{figure}[thbp]
\centering        
\vspace*{-0.2cm}
\includegraphics[width=0.45\textwidth,height=3.0in]{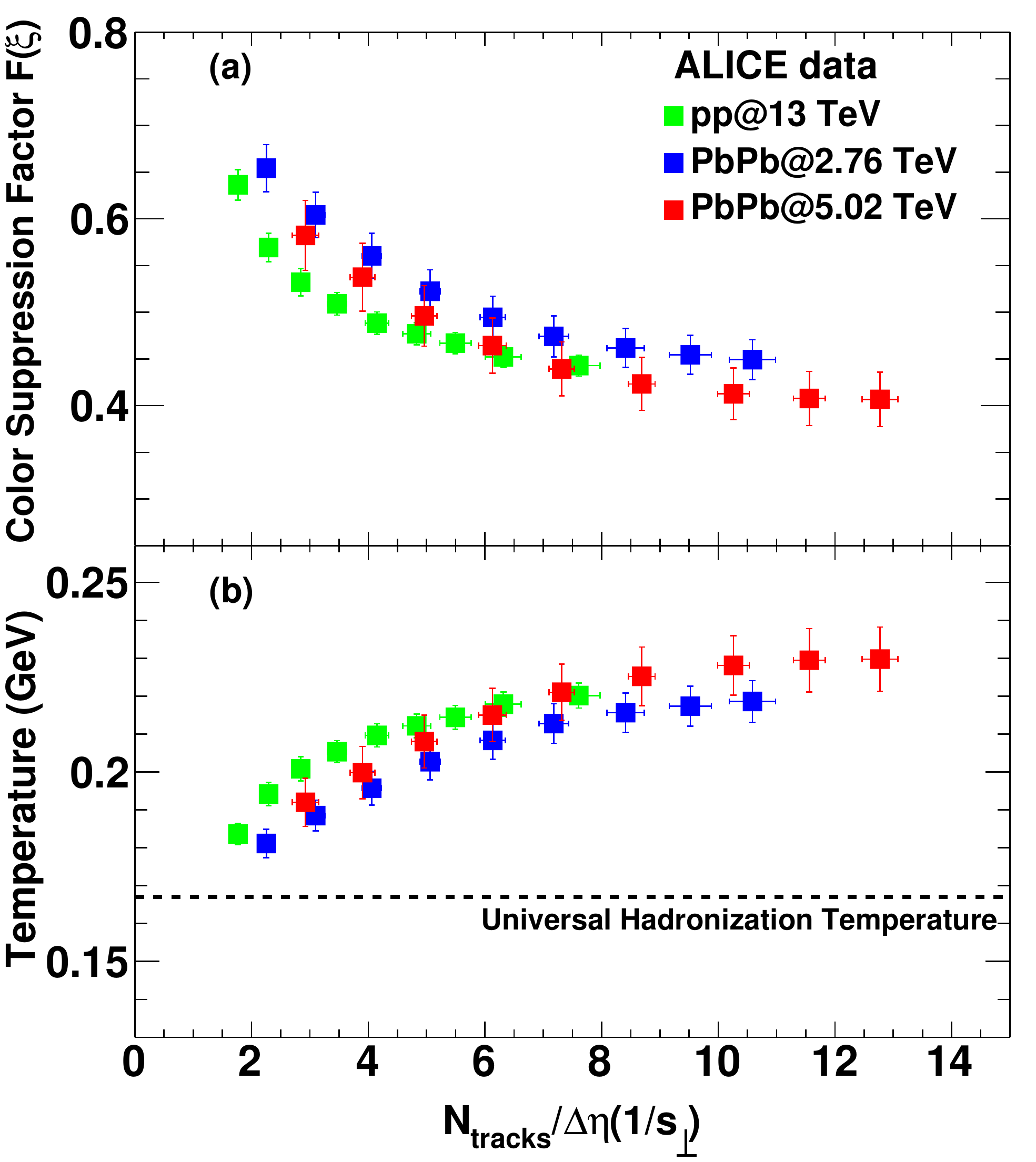}
\vspace*{-0.1cm}
\caption{ (a). Color Suppression Factor $F(\xi)$ in Pb-Pb  and ${\bf  pp}$ collisions vs  N$_{tracks}/\Delta\eta$ scaled by the transverse area $S_{\perp}$. $N_{tracks}$ is the  charged particle multiplicity in  the pseudorapidity range $|\eta| < $ 0.8 with $\Delta\eta$ = 1.6 units.
 For ${\bf pp}$ collisions $S_{\perp}$ is multiplicity dependent as obtained from IP-Glasma model \cite{cross}. In case of  Pb-Pb collisions the nuclear overlap area was obtained using the Glauber model \cite{glauber}. (b). Temperature vs N$_{tracks}/\Delta\eta$ scaled by  $S_{\perp}$ from  Pb-Pb and  ${\bf pp}$  collisions. The line  $\sim 165 $ MeV is the universal hadronization temperature \cite{bec1}. }
\label{fxi_temp}
\end{figure}
\section{Extraction of the color suppression factor  $F(\xi)$}
In the present work we have extracted $F(\xi)$ in
Pb-Pb collisions using ALICE data from the transverse momentum spectra of charged particles at $\sqrt{s_{NN}}$= 2.76 and 5.02 TeV at various centralities \cite{alicepb}. In case of ${\bf pp}$ collisions at $\sqrt s$ = 13 TeV,  $F(\xi)$ has been obtained in high multiplicity events  \cite{alicepp}. Only the softer sector of the spectra with $p_{t}$ in the range 0.15-1.0 GeV/c is considered. We have checked that the extension of the soft range up to $p_{t}$ = 2 GeV/c does not change significantly our results. 
To evaluate the initial value of $F(\xi)$ from data  a parameterization of the experimental data of $p_{t}$ distribution in low energy ${\bf pp}$ collisions at $\sqrt s$ = 200 GeV, where strings have low overlap probability, was used \cite{eos}.  The charged particle spectrum is described by a power-law \cite{review15}
\begin{equation}
  d^{2}N_{c}/dp_{t}^{2} = a/(p_{0}+p_{t})^{\alpha},
  \label{spectra1}
\end{equation}
where $a$ is the normalization factor, $p_{0}$ and $\alpha$ are fitting parameters with $p_{0}$= 1.98 and  $\alpha$ = 12.87 \cite{eos}. This parameterization is used both in Pb-Pb, and in  ${\bf pp}$ collisions to take into account the interactions of the strings \cite{review15}. The parameter $p_{0}$ in  Eq.~(\ref{spectra1}) is for independent strings and gets modified
%The color suppression factor $F(\xi)$ encodes the effect of the interaction among strings once they overlap.
%
\begin{equation}
p_{0} \rightarrow p_{0} \left(\frac {\langle nS_{1}/S_{n} \rangle^{mod}}{\langle nS_{1}/S_{n} \rangle_{pp}}\right)^{1/4}, 
\end{equation}
In  ${\bf pp}$ collisions at low energies only two strings are exchanged with low probability of interactions, so that
$\langle nS_{1}/S_{n} \rangle_{pp}$ $\approx$ 1, which transforms Eq.~(\ref{spectra1}) into
\begin{equation} 
  \frac{d^{2}N_{c}}{dp_{t}^{2}} = \frac{a}{(p_{0} \sqrt {1/F(\xi)^{mod}}+{p_{t}})^{\alpha}},
  \label{spectra2} 
\end{equation}
where $F(\xi)^{mod}$ is the modified color suppression factor and is used in extracting $F(\xi)$  both in Pb-Pb and ${\bf pp}$ in high multiplicity events. The color suppression factor $F(\xi)$ encodes the effect of the interaction among strings once they overlap.
 In the thermodynamic limit $F(\xi)$ is related to the
string density $\xi $ \cite{review15}
\begin{equation}
   F(\xi) = \sqrt {\frac {1-e^{-\xi}}{\xi}}.
  \label{fxitoxi}
\end{equation}
\begin{figure}[thbp]
\centering        
\vspace*{1.0cm}
\includegraphics[width=0.5\textwidth,height=3.0in]{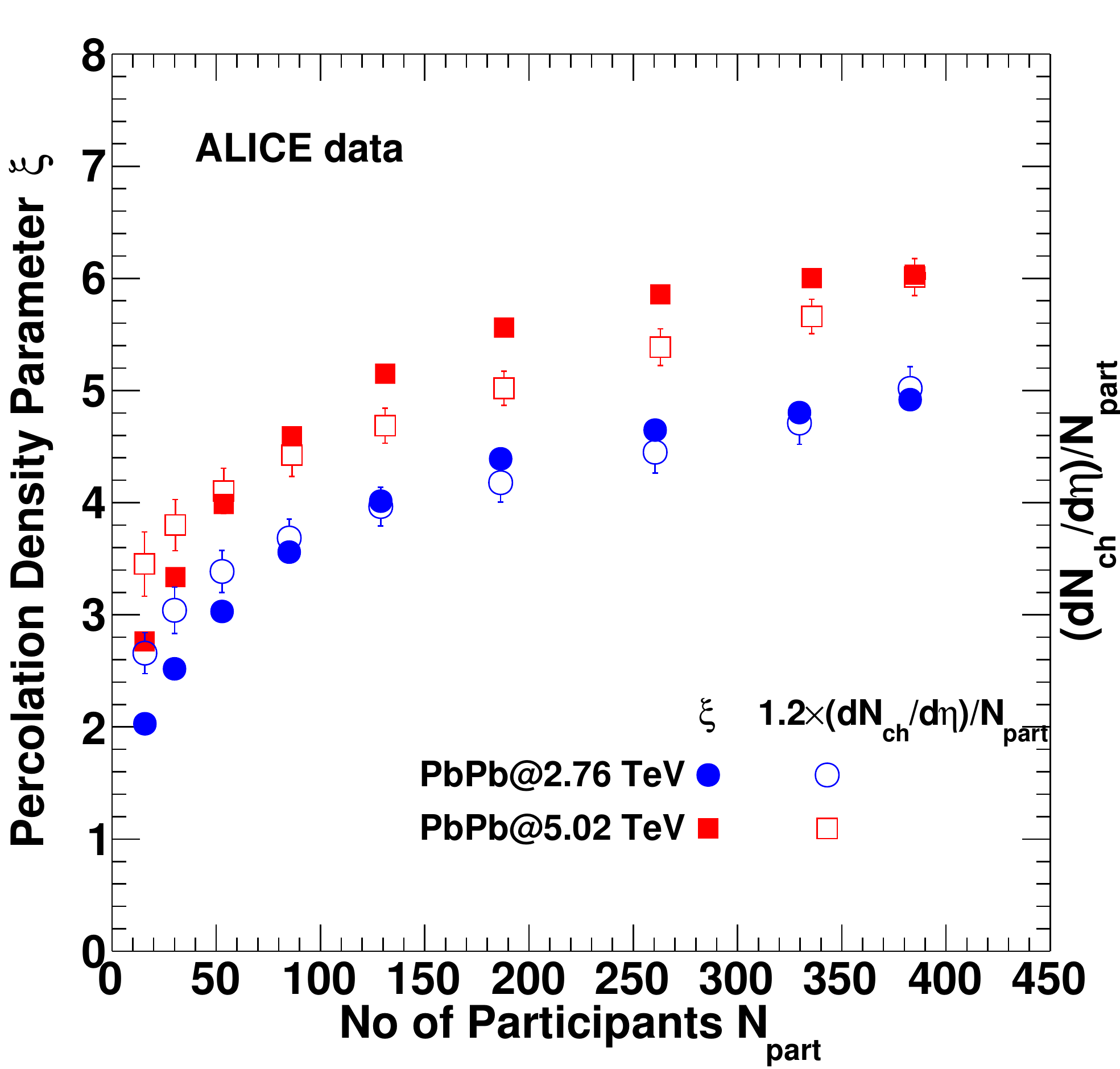}
\vspace*{-0.2cm}
\caption{ $\xi$ as a function of N$_{part}$ for  Pb-Pb collisions at  $\sqrt{s_{NN}}$ = 2.76 and 5.02 TeV.
  (dN$_{ch}$/d$\eta)/N_{part}$ is also  shown on the figure for comparison with $\xi$.}
\label{ximultnpart}
\end{figure}
The partons produced in the decay of a cluster interact with the color of the clusters in their way out the collision surface, modifying their momenta. This reproduces the experimental harmonics of the azimuthal asymmetry observed in ${\bf pp}$, ${\bf p-A}$, and A-A collisions. The size of the momentum modification is small and do not change the $p_{t}$ distribution. \cite{flow1,flow2,flow3,flow4}.
\section{Results and discussions}
Figure~\ref{fxi_temp}(a) shows  $F(\xi)$ as a function of  N$_{tracks}/\Delta\eta$ scaled by the transverse area $S_{\perp}$ for both Pb-Pb and ${\bf pp}$ collisions.
The error on $F(\xi)$ is $\sim$ 3$\%$ and is obtained with changing the fitting range of the transverse momentum spectra. 
For ${\bf pp}$ collisions $S_{\perp}$ is multiplicity dependent according to the IP-Glasma model \cite{cross,cross1,cross2}. In the case of Pb-Pb collisions the nuclear overlap area was obtained using the Glauber model \cite{glauber}.
A universal scaling behavior is observed in hadron-hadron and nucleus-nucleus collisions.
The percolation density parameter $\xi$ is obtained from $F(\xi)$ using the relation Eq.~(\ref{fxitoxi}) is shown in Fig.~\ref{ximultnpart} as a function of number of participants $N_{part}$ for  Pb-Pb collisions at  $\sqrt{s_{NN}}$ = 2.76 and 5.02 TeV. It is observed that $\xi$ rises slowly at higher $N_{npart}$. This behavior is similar to measured (dN$_{ch}$/d$\eta )/N_{part}$ as shown in  Fig.~\ref{ximultnpart}.
%\section{temperature}

The connection between $F(\xi)$ and the temperature $T(\xi)$ involves the Schwinger  QED$_{2}$ mechanism (SM) for particle production. 
The Schwinger distribution for mass less particles is expressed in terms of $p_{t}^{2}$ \cite{review15,pajares3}
\begin{equation}
  dn/d{p_{t}^{2}} \sim e^{-\pi p_{t}^{2}/x^{2}}
  \label{qed}
\end{equation}
where the average value of the string tension is  $\langle x^{2} \rangle$.
The tension of the macroscopic cluster fluctuates around its mean value because the chromoelectric field is not constant. The origin of the string fluctuation is related to the stochastic nature of the QCD vacuum.  Since the average value of the color field strength
must vanish, it cannot be constant but changes randomly from point to point \cite{bialas,dosch}.
Such fluctuations lead to a Gaussian distribution of the string tension and transforms SM into the thermal distribution \cite{bialas}
 \begin{equation}
\frac{dn}{dp_{t}^{2}} \sim \exp \left (-p_{t} \sqrt {\frac {2\pi}{\langle x^{2} \rangle}} \right ) ,
\label{bia}
\end{equation}
  with $\langle x^{2} \rangle$ = $\pi \langle p_{t}^{2} \rangle_{1}/F(\xi)$.
The temperature is expressed as \cite{eos,pajares3}  
\begin{equation}
T(\xi) =  {\sqrt {\frac {\langle p_{t}^{2}\rangle_{1}}{ 2 F(\xi)}}}.
\label{temp}
\end{equation}
\begin{figure}[thbp]
\centering        
\vspace*{0.0cm}
\includegraphics[width=0.50\textwidth,height=3.0in]{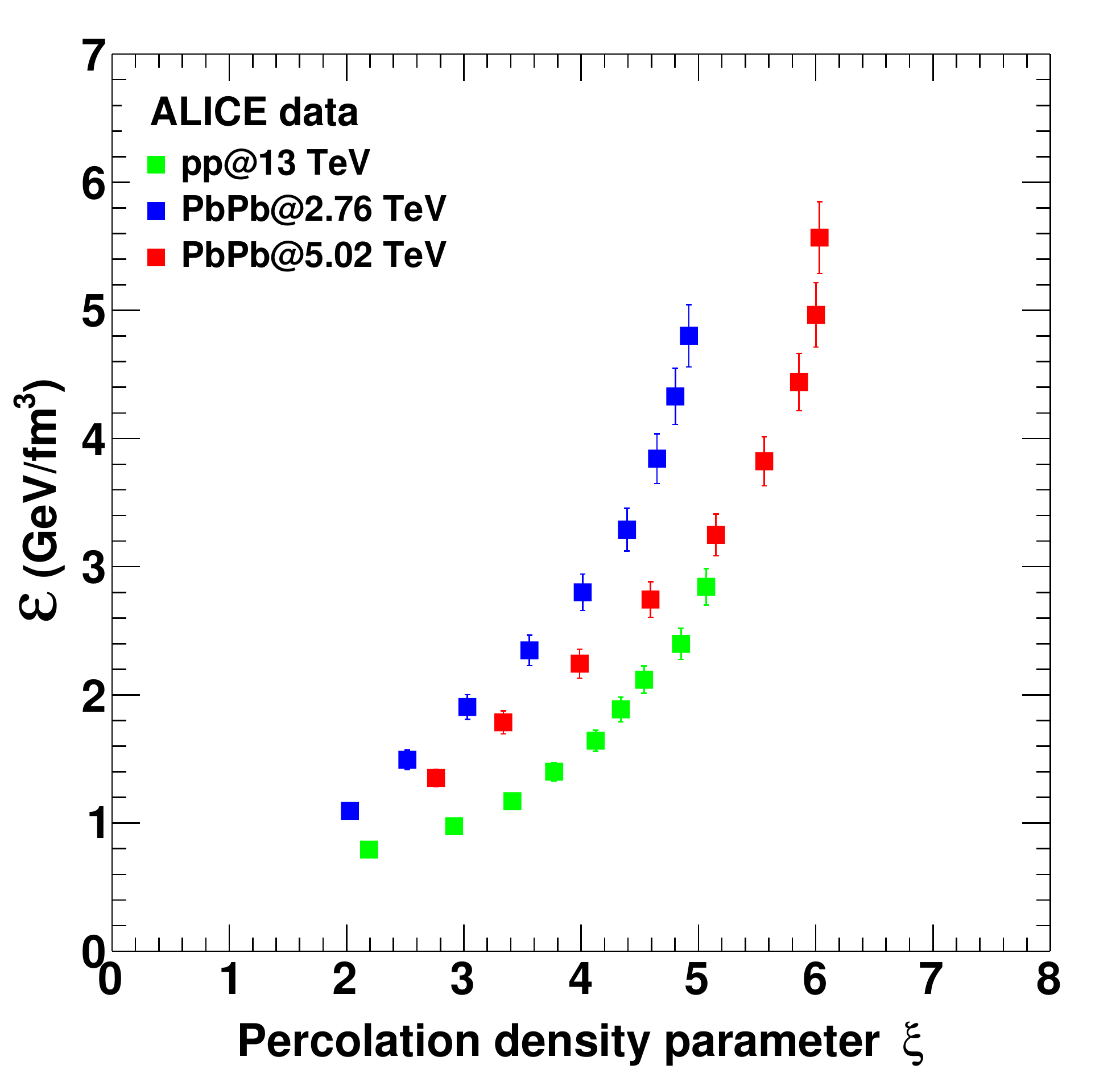}
\vspace*{-0.3cm}
\caption{Energy density ${\bm \varepsilon}$ (GeV/$fm^{3}$) as a function of string density $\xi$ for Pb-Pb and ${\bf pp}$ collisions at LHC energies. The energy density is obtained using Eq.~(\ref{bjk}) and $\xi$ is obtained using Eq.~(\ref{fxitoxi}).}
  \label{energy}
\end{figure}
We adopt the point of view that the universal hadronization
temperature is a good measure of the upper end of the cross over phase transition temperature $T_{h}$ \cite{bec1}. The single string average transverse momentum  ${\langle p_{t}^{2}\rangle_{1}}$ is calculated at $\xi_{c}$ = 1.2 with the  universal hadronization temperature 
$T_{h}= $ 167.7 $\pm$ 2.6 MeV~\cite{bec1}. This gives \mbox{$ \sqrt {\langle {p_{t}^{2}} \rangle _{1}}$  =  207.2 $\pm$ 3.3 MeV}.

Figure \ref{fxi_temp}(b) shows a plot of temperature as a function of N$_{tracks}/\Delta\eta$ scaled by $S_{\perp}$.
N$_{tracks}$ is the  charged particle multiplicity in  the pseudorapidity range $|\eta| < $ 0.8 with $\Delta\eta$ = 1.6 units.
The temperature dependence, for the systems investigated, falls on a universal curve as a function of the scaled multiplicity.
%Temperature from both hadron-hadron and nucleus-nucleus  collisions fall on a universal curve when multiplicity is scaled by the transverse interaction area.
The horizontal line at $\sim $ 165 MeV is the universal hadronization temperature obtained from the systematic comparison of the statistical thermal model parametrization of hadron abundances measured in high energy  ${\bf e^{+}e^{-}}$,
${\bf pp}$, and A-A collisions \cite{bec1}. In Fig.\ref{fxi_temp}(b) the temperatures obtained for Pb-Pb and
${\bf pp}$ are above the hadronization temperature indicating deconfinemnet.

Recently, it has been suggested that fast thermalization in A-A and ${\bf pp}$ collisions can occur through the existence of an event horizon due to a rapid deceleration of the colliding nuclei \cite{casto12,alex}. The thermalization in this case is due to the Hawking-Unruh effect \cite{casto12,satz,hawk,unru}. In CSPM the strong color field inside the large cluster produces deceleration of the primary $q\bar{q}$ pair which can be seen as a thermal temperature by means of Hawking-Unruh effect \cite{hawk,unru}.  
%Barrier penetration of the event horizon leads to a partial loss of information and is the reason for the stochastic thermalization of the $q\bar{q}$ pairs.
In the CSPM the initial local temperature  determined at the string level becomes the global temperature of the initial state as the cluster spans the whole area above  the percolation threshold.

The QGP according to CSPM is born in thermal equilibrium  because the initial temperature is determined at the string level. Above the critical temperature  T$ >$ T$_{c}$ the  CSPM energy expands according to the Bjorken boost invariant 1D hydrodynamics \cite{bjorken}
\begin{equation}
{\bm \varepsilon }= \frac {3}{2}\frac { {\frac {dN_{c}}{dy}}\langle m_{t}\rangle}{S_{\perp} \tau_{pro}},
\label{bjk}
\end{equation}
where ${\bm \varepsilon}$ is the energy density, $S_{\perp}$ the nuclear overlap area, $m_{t}$  is the transverse mass and
$\tau_{pro}$ is the Schwinger production time for a boson (gluon)  ${\tau_{pro}} = 2.45/\langle m_{t}\rangle$  \cite{schw}.
Figure~\ref{energy} shows ${\bm \varepsilon}$ as a function of ${\xi}$  for Pb-Pb and ${\bf pp}$ collisions.
We observe a slow rise of  ${\varepsilon}$  for low values of ${\xi}$ followed by a faster rise later.
  This  is due to the nonlinear increase in multiplicity at higher ${\xi}$ values. In the case of Pb-Pb these data show a large departure of the scaling of the multiplicity per participant. This change occurs at
  ${\xi} \sim$  4.2 and $\sim $5.0 for Pb-Pb at 2.76 TeV and 5.02 TeV respectively. For ${\bf pp}$ the jump in ${\bm \varepsilon}$ is at ${\xi }\sim $  3.8.

\begin{figure}[thbp]
\centering        
\vspace*{-0.2cm}
\includegraphics[width=0.50\textwidth,height=3.2in]{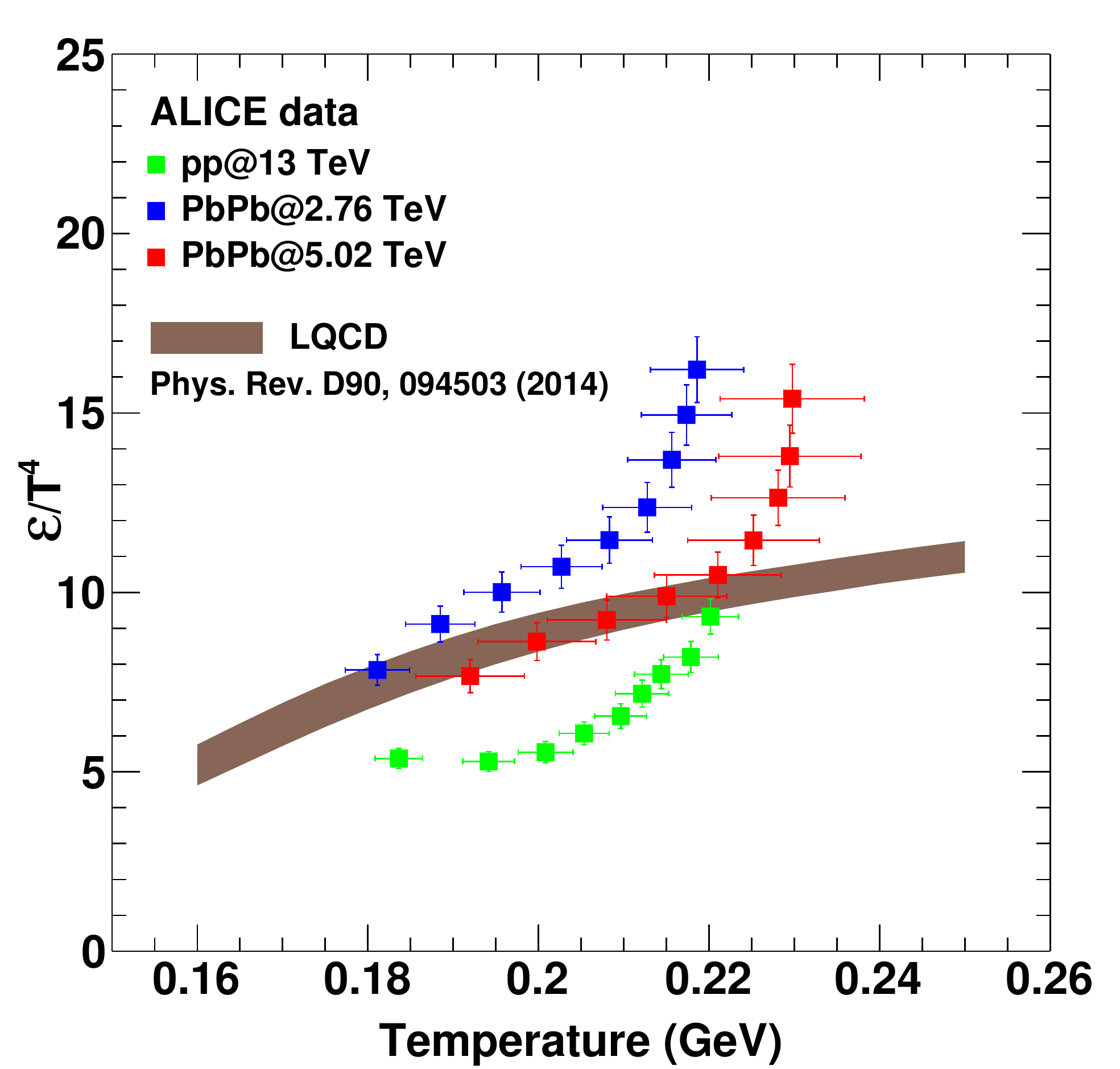}
\vspace*{-0.5cm}
\caption{ Dimensionless quantity ${\bm \varepsilon}/T^{4}$ as a function of temperature for Pb-Pb collisions at $\sqrt {s_{NN}}$ = 2.76 and 5.02 TeV. ${\bf pp}$ at $\sqrt {s}$ = 13 TeV is shown as  green rectangle. The brown band corresponds to LQCD simulations \cite{lattice14}. }
\label{et4}
\end{figure}

 In Fig.~\ref{et4} we show the results of dimensionless quantity ${\bm \varepsilon}/T^{4}$ for Pb-Pb collisions at two different energies and those for ${\bf pp}$ at 13 TeV. The results are compared with LQCD predictions for 2+1 flavors \cite{lattice14,wuppe14}.
It is observed that CSPM results agree with LQCD results up to the temperature of T $\sim$  210 MeV for the Pb-Pb collisions. Beyond this temperature the ${\bm \varepsilon/T^{4}}$ in CSPM rises much faster and reaches the ideal gas value of  ${\bm \varepsilon}/T^{4}$ $\sim$ 16 at T $\sim$ 230 MeV. In this region, there is a strong screening due to the large degree of overlapping of the strings, producing a faster approach to the quark gluon gas limit.

The DOF are obtained using the relation ${\bm \varepsilon}/T^{4} $= DOF ${\pi^{2}}/30$ \cite{wong}.
At T  $\sim ~ $ 210 MeV, ${\bm \varepsilon}/T^{4}$ $\sim$ 11 which corresponds to $\sim$ 33 DOF while at T  $\sim$ 230 MeV there are $\sim$ 47 DOF.  It is observed that Pb-Pb at $\sqrt {s_{NN}}$ = 2.76  TeV has similar features as seen at 5.02 TeV. In ${\bf pp}$ collisions at  $\sqrt {s}$ = 13 TeV only $\sim$ 33 DOF are reached. Our results are in agreement with the conclusions obtained studying the trace anomaly in a quasi particle gluonic model \cite{casto1,casto2}. In this model the DOF of the free gluons are also obtained for T $\simeq$ 1.3T$_{c}$ (T$_{c} \approx$ 165 MeV).
% lattice discussion

There are several reasons that can explain the disagreement between CSPM and lattice QCD for $T > 1.3 T_{c}$. First in A-A and pp collisions a strong magnetic field B is produced, which is larger for A-A than for pp and increases with energy.
The lattice studies of the chiral phase structure of three flavor QCD in a background magnetic field show that chiral condensate and the phase transition temperature always increases with B. The transition becomes stronger and turns into a first order instead of crossover \cite{mag2}. As B is higher in PbPb than in pp, we expect a higher phase transition temperature in these cases. 

Recently, it has been pointed out  using lattice simulations that in addition of the standard crossover phase transition at T $\sim$ 155 MeV, the existence of a new infrared phase transition at temperature T, 200 $<T_{IR}<$ 250 MeV. In this phase, asymptotic freedom works and therefore there is no interaction. In between these two temperatures there is coexistence of the short and long distance scales \cite{newphase}, which supports the present observation in our work.

A new phase in QCD also has been proposed studying the Dirac operator. While confining chromo-electric interaction is distributed among all modes of Dirac operator, the chromo-magnetic interaction is located predominantly in the near zero modes. Above T $\sim$ 155 MeV the near zero modes are suppressed but not the rest of the modes, surviving the chromo-electric interaction which is suppressed at higher temperature \cite{glozman}.

 \section{Conclusion}
We have used the Color String Percolation Model (CSPM) to explore the initial stage of high energy nucleus-nucleus and nucleon-nucleon collisions and determined the thermalized initial temperature of the hot nuclear matter at an initial time $\sim 1$ fm/c.
 For the first time the temperature and the energy density of the hot nuclear matter, from the  measured charged particle spectra using ALICE data for Pb-Pb collisions at $\sqrt {s_{NN}}$ = 2.76 and 5.02 TeV and ${\bf pp}$ collisions at $\sqrt {s}$ = 13, TeV have been obtained. 
The dimensionless quantity ${\bm \varepsilon}/T^{4}$ is evaluated to obtain the number of degrees of freedom (DOF) of the deconfined phase.
We observe two features hitherto not reported: the existence of two temperature ranges in the behavior of the Pb-Pb system DOF,
and a clear departure from the LQCD results regarding the maximum number of DOF, which reaches values in agreement with the Stephan Boltzmann limit for an ideal gas of quarks and gluons.

\begin{acknowledgements}
A. M. acknowledges the post-doctoral fellowship of DGAPA UNAM. Partial support was received by DGAPA-PAPIIT IN109817 and CONACYT A1-S-16215 projects. A. M. also thanks the Hungarian  National  Research,  Development  and Innovation Office (NKFIH) under the contract numbers OTKA K120660, NKFIH  2019-2.1.11-T ET-2019-00078, and 2019-2.1.11-T ET-2019-00050.
C. P. thanks the grant Maria de Maeztu Unit of excellence MDM-2016-0682 of Spain, the support of Xunta de Galicia under the project  ED431C 2017 and  project FPA 2017-83814 of Ministerio de Ciencia e Innovacion of Spain and FEDER. GP acknowledges the DGAPA sabbatical fellowship.  
\end{acknowledgements}

\end{document}